A Review of Exoplanetary Biosignatures

John Lee Grenfell

Department of Extrasolar Planets and Atmospheres (EPA), Institute of Planetary Research,

German Aerospace Centre (DLR), Rutherford Str. 2, 12489 Berlin, Germany

Email: lee.grenfell@dlr.de

Tel: +49 30 67055-9734

FAX: +49 30 67055-384

**Abstract**. *We review the field of exoplanetary biosignatures with a main focus upon atmospheric gas-phase species. Due to the paucity of data in Earth-like planetary atmospheres a common approach is to extrapolate knowledge from the Solar System and Early Earth to Earth-like exoplanets. We therefore review the main processes (e.g. atmospheric photochemistry and transport) affecting the most commonly-considered species (e.g. $O_2$, $O_3$, $N_2O$, $CH_4$ etc.) in the context of the modern Earth, Early Earth, the Solar System and Earth-like exoplanets. We consider thereby known abiotic sources for these species in the Solar System and beyond. We also discuss detectability issues related to atmospheric biosignature spectra such as band strength and uniqueness. Finally, we summarize current space agency roadmaps related to biosignature science in an exoplanet context.*

# 1. Introduction

The science of biosignatures i.e. signals which suggest the presence of life, is a wide and interdisciplinary field which is currently undergoing rapid expansion due in part to advances in exoplanet science. In this context, it is therefore convenient to split current biosignature methods into two groups, namely 'in-situ' and 'remote'. We define here in-situ biosignatures to include the study of fossil morphology and the direct examination of complex biotic molecules such as DNA and lipidic acids (see e.g. Horneck et al., [1] for a review). In-situ biosignatures are not our main focus here. We define remote biosignatures to refer to spectroscopic methods for detecting e.g. atmospheric species' abundance, (surface) reflectivity etc. In recent years there has been some cross-over from in-situ to remote methods e.g. for biosignatures involving isotopic ratios, surface pigments etc. as we will discuss.

This paper reviews the field of remote biosignatures in an exoplanetary context with a main focus on atmospheric species. Due to limited data in Earth-like exoplanetary science, a common approach when studying potential exoplanetary biosignatures is to consider what can be learned from the modern Earth, Early Earth and the Solar System. Based on these, one then extrapolates knowledge gained (together with theoretical model studies) to the study of Earth-like exoplanets.

Literature reviews which cover subjects related to exoplanetary biosignatures but which are not discussed in detail in this paper are cited throughout the manuscript. For conciseness however, we list here briefly some of the main review papers relevant to our subject. A review of conditions affecting potential habitability and biosignatures beyond the Solar System is provided in the Special Issue "Planetary Habitability and Life" edited by Spohn et al. [2] and references therein. The potential habitability of terrestrial planets orbiting in the Habitable Zone of M-dwarf stars (important targets in exoplanet science) was reviewed in Scalo et al. [3] and more recently by Shields et al. [4]. These are central objects in exoplanetary science and are targets of interest for forthcoming space missions. This is because firstly, M-dwarf stars are numerous in the solar neighborhood; secondly, planets orbiting smaller, fainter stars have stronger (planet/star) contrast ratio signals and thirdly a fainter star suggests a closer HZ where planets have faster orbits meaning more rapid collection of data. There are however possible caveats regarding the habitability of these objects such as possible planetary tidal-locking due to the close proximity of the star. This could lead to slower planetary rotation and a possible weakening in the planet's magnetosphere hence potentially strong bombardment of the atmosphere with high energy particles as detailed in the Scalo and Shields reviews mentioned above. The extent to which these processes could be related however is generally not well-constrained. Lammer et al. [5] reviewed evolution of planetary environments towards habitable conditions. Regarding biosignatures, the recent NASA NExSS (**N**exus for **Ex**oplanet **S**ystem **S**cience) initiative has produced comprehensive reviews which discuss finding, assessing, detecting and planning observation programs for exoplanetary biosignatures, as detailed in Schwieterman et al., [6]; Meadows et al., [7]; Catling et al., [8]; Walker et al., [9]; and Fujii et al., [10]. Reviews by Seager [11, 12] summarize responses of gas-phase species in biosignature science. A review of biosignature-related photochemical responses on modern Earth is provided by e.g. Wayne [13]. Reviews of potential biosignature species on Mars (and Venus) are provided by e.g. Krasnopolsky [14] and Atreya et al [15].

The paper is organized as follows. Section 1 discusses current definitions of life (1.1) and the steps involved when searching for life remotely (1.2). Section 2 discusses biosignatures for vegetation and biological pigments. Section 3 deals with spectrophotometry of gas-phase species which have been suggested as potential biosignatures. Section 4 summarizes the application of spectropolarimetry in remote biosignature science. Section 5 discusses reaction network properties in a biosignature context. Section 6 summarizes studies of Earth-like planets without life ("dead Earths"). Section 7 discusses current space agency roadmaps in astrobiology related to searching for remote biosignatures. Section 8 provides a brief summary.

## 1. 1 Defining Life

There is currently no universally-accepted definition of life. Frequently cited [16] is the NASA definition of a "self-sustaining chemical system capable of Darwinian evolution". Cleland and Chyba [17] review the challenges in defining life and its needs.

## 1.2. Searching for Life Remotely

Assessing remote biosignatures is a multi-tiered process. The first step involves finding suitable targets for study. This requires prior information on the fraction of stars which host potentially Earth-like planets - the "eta Earth" ($\eta_{earth}$) parameter as discussed in e.g. Kane et al. [18]. The second step involves obtaining the measurement itself. Common methods discussed in the context of biosignature studies include transit spectrophotometry - for which there exists a large literature (see e.g. Rauer et al., [19]) and spectropolarimetry (see e.g. Stam [20]; Sterzik et al., [21]; see below). Deriving the atmospheric properties such as temperature, pressure and composition from the observed spectra usually involves dealing with degeneracies which may arise during the retrieval process. These can be addressed e.g. in the case of atmospheric spectroscopy by combining information from e.g. absorption band strength, shape and spectral slope as discussed in Benneke and Seager [22]. The third step involves a process of elimination to ensure that the biosignature candidate is really attributable to life. This requires characterization of the planetary environment and evolution in order to discount any abiotic sources which could be falsely interpreted as life. There now follows an overview of the various proposed biosignature approaches.

## 2. Vegetation and biological pigments

The 'red edge' refers to the ~five times or more increase in reflectance at ~(700-750)nm displayed by vegetation. It arises due to the strong contrast between the absorption of chlorophyll and the otherwise highly reflective leaf. The strong reflectivity arises due to stacked layers of plant cells with gaps in-between which leads to efficient multiple scattering (Seager et al., [23]). For an Earth-like planet this study suggested that the global red edge signal amounts to only a few percent in reflectivity. Such a signal could nevertheless be detectable for bright targets using future proposed biosignature space missions. The same study also suggested searching for seasonally-varying changes in the 'red-edge' since detecting this could add confidence that the signal is a signature of life. Other studies e.g. Pallè et al. [24] estimated the Earth's red edge signal from Earthshine during lunar eclipse. For a detailed discussion see Kaltenegger et al. [25] and references therein. Analogous to the 'red edge', the 'purple edge' was proposed for early Earth atmospheres (Sanroma et al., [26]) due to purple bacteria which could have thrived under Archaean conditions. Schwieterman et al. [27] extended the 'red edge' principle by considering non-photosynthetic pigments as exoplanetary biosignatures.

## 3. Atmospheric Spectrophotometry of Gas-Phase Species

Earth's atmosphere and life on our planet have co-evolved in numerous, intricate ways. The atmosphere influences the planetary climate, protects the surface from harmful radiation or/and cosmic rays and enables the prolonged existence of liquid water required by all life. Kasting and Catling [28]; de Vera and Seckbach [29] and Cockell et al. [30] provide excellent reviews of habitability - the potential to host life - including the role played by atmospheres. Life has also modified Earth's atmosphere, imprinting its existence there in the form of so-called atmospheric biosignatures.  In this section we review current knowledge of atmospheric biosignature gases such as molecular oxygen ($O_2$), ozone ($O_3$), nitrous oxide ($N_2O$) etc. and their associated spectrophotometric signals for planets in the (Classical) Habitable Zone (HZ) (Kasting et al., [31]) which is defined as the annulus region around a star where a planet can maintain liquid water. The field of atmospheric biosignatures is rapidly expanding so the material presented here is not exhaustive but instead provides a brief overview. Since the subject is interdisciplinary, we explain some standard (subject-dependent) concepts where we feel necessary.

Seager et al. [11, 12] reviewed atmospheric biosignatures. They categorized them into those produced by metabolism and those produced by stress or signaling. They noted that the Earth features a large variety of small, volatile compounds which have both biotic and non-biotic sources. Seager et al. [11] also applied a model to estimate biomass amounts as a plausibility check when considering biomass emissions on Earth-like planets. An example of an atmospheric biosignature is Earth's oxygen-rich atmosphere which arises mainly due to photosynthetic activity. A central question is whether such features could be detected spectroscopically in order to infer unambiguously the presence of life in an exoplanetary context.

This section reviews current understanding of atmospheric gas-phase biosignatures with a focus on atmospheric species' budgets (i.e. their global sources and sinks) and their spectral responses.  The section is primarily structured into sub-sections according to biosignature gas-phase species. Due to the paucity of data for Earth-like exoplanets, most studies are based upon theory and upon what is known from the (Early) Earth and Solar System planets. Each atmospheric species is therefore discussed in the context of modern Earth, Early Earth, the Solar System and Earth-like exoplanets and concludes with "biosignature aspects" which discusses detection-related issues such as the strength and uniqueness of its spectral features. Section 3 closes with a brief discussion of additional, proposed atmospheric biosignatures, including technosignatures (3.5.4), redox disequilibrium (3.5.5) and isotopic ratios (3.5.6).

### 3.1 Molecular Oxygen ($O_2$)
### 3.1.1 Modern Earth

Earth's present-day atmosphere features an $O_2(g)$ volume mixing ratio (vmr), $\xi_{o2}$=0.21 which corresponds to a total mass of ~$1.2 \times 10^{21}$g. $O_2(g)$ is chemically a rather inert species with a long atmospheric lifetime, $\tau_{O2}$ (where $\tau$ is defined as the time for the concentration to fall due to atmospheric sinks alone by a factor equal to the base of the natural logarithm ['e']) of the order of several thousand years (Lasaga and Ohmoto, [32]). $O_2(g)$ features a constant vmr in Earth's atmosphere up to ~80km altitude (Brasseur and Solomon, [33]). At higher levels $O_2$(g) is photolysed into oxygen atoms, the three-body recombination of which generates the so-called oxygen airglow, a feature also observed in the atmospheres of Mars and Venus (see Slanger and Copeland, [34] for a review).

**Sources** – the main source of atmospheric $O_2$(g) occurs via photosynthesis (P). Cyanobacteria harvest the energy of sunlight in the presence of chlorophyll and nutrients as represented by the general equation: $CO_2$+$H_2O$→"$CH_2O$"+$O_2$. The reverse process, respiration (R) and decay of organic material involves consumption of "$CH_2O$" (organic matter) with $O_2$(g) to produce energy which drives metabolic processes. Over timescales of up to a few million years the rates of P and R are approximately equal and opposite i.e. similar amounts of $O_2$(g) are produced via P as consumed via R. How then, can these processes produce a net $O_2$ (g) source which has led to the formation of Earth's oxygen-rich atmosphere? Over timescales longer than several millions of years, organic matter produced from P gathers as sediment on the ocean floor where it is gradually compressed and subducted into the Earth's mantle in a process termed "burial" (Holland, [35]) which is favored by plate tectonics. Burial leads to a removal of "$CH_2O$" which would otherwise be available for R, hence results in a net excess in $O_2$(g) which enters the atmosphere. Lasaga and Ohmoto [32] estimate a source of ~320Tg/year $O_2$(g) via this mechanism.

A second, weaker source for $O_2$(g) on the modern Earth arises via atmospheric escape. Water can photolyse in Earth's stratosphere and above which results overall in the process: $2H_2O$+hv→$O_2$+4H. If the resulting H-atoms escape into space then this mechanism constitutes a net source of $O_2$(g). This process however, likely proceeds on modern Earth at a rate about two orders of magnitude slower than the $O_2$(g) source from burial (Yung and DeMore, [36]. On early Earth however, where the early Sun was much more active in EUV, this process led to rapid loss of the primary atmosphere's thick hydrogen envelope. Other abiotic sources for $O_2$(g) have been proposed in the context of exoplanets (see above) but these are expected to be  very weak on the modern Earth.

**Sinks** – there are two main sinks for $O_2$(g) in Earth's modern atmosphere. The first involves reduced atmospheric gases derived from volcanic outputs (Catling and Claire, [37]) and makes up (80-90%) of the total $O_2$(g) atmospheric removal on modern Earth. The second sink which makes up the remaining (10-20%) occurs via weathering on the Earth's surface (e.g. Holland, [38]).

**Model studies** - Overall, sources and sinks for $O_2$(g) are approximately in balance in modern Earth's atmosphere. There are numerous model studies of the $O_2$(g) budget (e.g. Holland, [39]; Kump, [40]; Van Capellen and Ingall, [41]; Lenton and Watson, [42]; Berner et al., [43] and Berner, [44]). Gebauer et al. [45] analysed chemical pathways affecting $O_2$(g) on the modern and Early Earth (assuming as model input an $O_2$ surface evolution with time based on proxy data). Results suggest complex oxidation pathways destroy $O_2$(g) on the lower levels whereas $O_2$(g) can be formed abiotically via e.g. $CO_2$ photolysis on the upper levels.

### 3.1.2 Early Earth

Understanding The evolution of $O_2$(g) in Earth's atmosphere is a central focus in Earth (e.g. Kump [46]). An initial rise termed the "Great Oxidation Event" (GOE) occurred (2.4 to 2.7) Gyr ago followed by a smaller, second increase termed the "Second Oxidation Event" (SOE) at around 0.6Gyr ago. The causal mechanism(s) behind these two events is/are still debated. Suggested mechanisms include e.g. a stronger burial rate (possibly due to changes in continental breakup) (Kump et al., [47]) or, a lowering in the reducing nature of volcanic emissions which enabled $O_2(g)$ to rise (Gaillard et al., [48]). A wide range of geological and atmospheric proxies support the shape of the evolutionary track of the GOE. See e.g. Holland [49] for an overview. Lyons et al. [50] examined trace elements such as molybdenum in ancient marine sediments and suggested a weaker GOE rise in $O_2(g)$ of up to ~$10^{-3}$ PAL i.e. about one hundred times lower than previously estimated [46].

The $O_2(g)$ evolution is likely linked with the climate history of Earth. For example, the GOE is proposed to have led to a reduction in atmospheric methane ($CH_4$) (a strong greenhouse gas) since rising $O_2(g)$ likely poisoned $CH_4$-producing (methanogenic) bacteria which consequently could have triggered a catastrophic global ice age (the "Snowball Earth") (see e.g. Kasting, [49]). The "Faint Young Sun" (FYS) paradox refers to the inconsistency between the morphology of ancient soil (paleosols) (whose flow-shaped appearance suggests liquid water on the surface of the Early Earth) and the climate predictions of atmospheric models which suggest a frozen surface. Feulner et al. [51] provide a review of mechanisms suggested to address the FYS paradox. Clearly, solving this issue would also further our understanding of the evolution of $O_2(g)$. Recent proxy data is currently revising our understanding of $O_2(g)$ on Early Earth e.g. molybdenum isotope data (Anbar et al., [52]) suggest small, localized "whiffs" of increased $O_2(g)$ prior to the GOE.

### 3.1.3 Solar System

Both Mars and Venus feature $CO_2$-dominated atmospheres which can form small amounts of $O_2(g)$ abiotically via: $CO_2+hv \rightarrow CO+O$ followed by: $O+O+M \rightarrow O_2+M$ where 'M' denotes any gas-phase third-body required to remove excess vibrational energy. The cold, thin Martian atmosphere features $\xi_{o2}$=1.4x$10^{-3}$ vmr whereas the hot, thick Venusian atmosphere features $\xi_{o2}$=3x$10^{-7}$ vmr (see Yung and DeMore [36] and references therein). These values are clearly much smaller than the biotically-produced $O_2(g)$ on Earth. It is interesting to note, though that the early Martian atmosphere may have contained the equivalent mass of oxygen as several tens of meters of water (Zhang et al., [53]) covering the surface, which was gradually lost via escape. Abiotic production of $O_2(g)$ in $CO_2$-dominated atmospheres is sensitive to potentially complex catalytic cycles involving e.g. hydrogen or nitrogen oxides which control the regeneration of CO back into $CO_2$ and which depend e.g. upon the UV-environment. Although abiotic production of $O_2(g)$ is weak on Mars and Venus it could be much stronger for certain Earth-like exoplanets. Energetic particles impinging on rocky or icy surfaces can lead to the formation of molecular $O_2(g)$. This species has been suggested to occur (although only at very small amounts) in the (exo)-atmosphere(s) of Mercury (Hunten et al., [54]), Europa (Hall et al., [55]), Ganymede (Hall et al., [56]) and Rhea (Teolis et al., [57]).

### 3.1.4 Earth-like Exoplanets

Owen [58] suggested to search spectroscopically for atmospheric $O_2(g)$ in order to detect extraterrestrial life. Although little is known about potential oxygen cycles on Earth-like exoplanets, model studies can give some indication as to how the sources and sinks of atmospheric oxygen could be affected on such worlds compared with the Earth.

**Sources** - Kiang et al. [59] discussed the potential for photosynthesis analogues occurring on Earth-like planets orbiting in the Habitable Zone (HZ) of M-dwarf stars where the incoming insolation peak is shifted to lower energies. These are central objects of study in exoplanet science as mentioned above (see e.g. reviews by Scalo et al., [3]; Shields et al., [4]). The Kiang study suggested that such planets would feature (10-50%) weaker photosynthetic activity than on Earth in the visible wavelength but with values potentially exceeding Earth's activity in the infra-red.

Burial on Earth (a strong source for $O_2(g)$ as already discussed) is more efficient in continental shelf regions (see e.g. Hartnett et al.,[60]) due to more efficient mixing leading to an enhanced abundance of nutrients. This suggests that Earth-like planets with extensive continental shelves could favour stronger burial hence a stronger biotic source of $O_2(g)$. Burial and subduction are however linked with active plate tectonics; the efficiency of these processes on Earth-like planets and Super-Earths is extensively debated (see e.g. Stamenković et al., [61]; Tackley et al., [62]; Noack and Breuer, [63]).

Regarding the abiotic atmospheric $O_2(g)$ source discussed above i.e. $H_2O(g)$ photolysis followed by H-escape as mentioned above - this process could be highly efficient for low mass planets in high EUV environments e.g. for planets orbiting pre-main sequence and early post-main sequence stars. Luger and Barnes [63] suggested that high levels of X-ray Ultra Violet (XUV) radiation from pre-main sequence M-dwarf stars could lead to rapid photolysis of $H_2O$ and strong escape of the resulting H in the planetary atmosphere.

$CO_2(g)$ photolysis (see above) could constitute a strong abiotic source for $O_2(g)$ under certain conditions. Tian et al. [65] suggested that Earth-like planets orbiting in the HZ of M-dwarf stars could feature up to 3 orders of magnitude more $O_2(g)$ via $CO_2$ photolysis than their counterparts orbiting FGK stars. Harman et al. [66] investigated this issue further noting the potential importance of global redox balance and ocean chemistry. Such strong abiotic $O_2(g)$ production arose due to the weak near-UV (NUV) output from the central star. Weak NUV weakens the release of OH from its reservoirs hence weakens the HOx-catalysed reaction: $CO+O\rightarrow CO_2$, favoring a rise in O hence a rise in $O_2$ via: $O+O+M\rightarrow O_2+M$. The model study by Domagal-Goldman et al. [67] on the other hand, calculated only modest abiotic $O_2$ amounts (several orders of magnitude lower than the Tian et al. study) for such a scenario. They suggested that their different result could have arisen due to different treatments of CO removal from the atmosphere related to a detailed treatment of reduced input from volcanos and mid-ocean ridges. The model study of Hu and Seager [68] suggested $O_2$-rich atmospheres could form abiotically for Super-Earths having elemental compositions with mole fractions $X_H<0.3$ and $X_C/X_O<0.5$. Wang et al. [69] suggested that such $O_2$-rich atmospheres formed from abiotic $CO_2(g)$ photolysis would also feature a strong CO(g) band. Their work therefore suggested CO(g) to be an "anti-biosignature." Schwieterman et al. [70,71] proposed ways of distinguishing abiotic $O_2$ signals spectroscopically.

**Sinks** – reaction with reducing gases (like $CH_4$) in the planet's atmosphere has been proposed to constitute an important atmospheric sink for $O_2(g)$. Model studies such as Segura et al. [72,73] suggested high levels of $CH_4(g)$ (around 100-1000 times those on the modern Earth) in the atmospheres of Earth-like planets orbiting M-dwarf stars. This is because weaker UV output from the star slows the reaction:

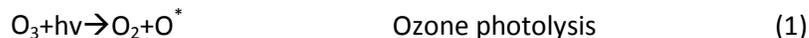

$$O_3 + h\nu \rightarrow O_2 + O^* \qquad\qquad \text{Ozone photolysis} \qquad\qquad (1)$$

In reaction (1) $O^*$ denotes electronically-excited oxygen atoms. A slowing in reaction (1) leads to a weakening in:

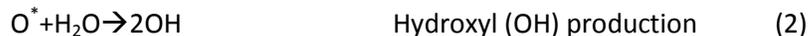

$$O^* + H_2O \rightarrow 2OH \qquad\qquad \text{Hydroxyl (OH) production} \qquad (2)$$

-which leads to a rise in $CH_4$ since OH is its main sink.

### 3.1.5 Earth-like Exoplanets in time

The Earth's atmosphere $O_2(g)$ budget has undergone major changes during its history as already discussed. Reducing uncertainty in the evolution of $O_2(g)$ during Earth's history could clearly impact the discussion on its detection in Earth-like atmospheres. Numerous factors including e.g. the stellar environment, the planetary orbital parameters etc. likely played a role in shaping this evolution and these same factors are expected to influence potential atmospheric biosignatures on Earth-like planets. Grenfell et al. [74] discussed co-evolution mechanisms for life, climate and biosignatures in the context of Early Earth and Earth-like planets. Schindler and Kasting [75] and Kaltenegger et al. [76] calculated the evolution of atmospheric biosignatures over geological epochs for an Earth-like planet.

### 3.1.6 Biosignature aspects

$O_2(g)$ features a rather weak absorption band in the visible region at 0.76 microns. Des Marais et al. [77] was an important study which discussed spectral features of Earth-like atmospheric biosignatures including $O_2(g)$. Rodler and López-Morales [78] suggested that the 0.76 micron band would need several tens of transits to be detectable at three-sigma with the next generation of ground-based telescopes for an Earth-analogue orbiting a bright M-dwarf star with spectral class greater than M3. Kawahara et al. [79] suggested to examine the near-IR 1.27 microns band since adaptive optics from the ground is more favorable in this wavelength region than in the visible. Their study suggested a 5σ detection (calculated against photon noise) for $O_2$ in this IR band assuming Earth analogues (if they exist) for up to 50 close-by stars assuming 5 hours exposure on a 30m telescope. Misra et al. [80] suggested absorption bands from the $O_2$-$O_2$ dimer in the IR as a spectral feature for detecting molecular oxygen. Snellen [81] discuss the challenges of $O_2(g)$ detection on Earth-like planets using the E-ELT. Loeb and Maoz [82] discuss the detectability of atmospheric $O_2(g)$ for on an Earth-like planet which has migrated into the HZ of a white dwarf. Their study claims detectability of the 0.76μm $O_2(g)$band with 5h total exposure time on the James Webb Space Telescope (JWST). In the longer-term, direct imaging of Earth-like planets around different main sequence stars is a central science goal of proposed space-based missions such as the Large Aperture UV-Optical-Infrared (LUVOIR) telescope and the Habitable Exoplanet (HabEx) Imager (see section 7).

### 3.1.7 Concluding remarks

The case of $O_2(g)$ illustrates the emerging consensus that atmospheric biosignature species should be assessed in the context of the planetary evolution and environment. This means knowledge of e.g. the stellar spectrum and the evolution history of the atmosphere in order to rule out the false positives (abiotic signals which mimic life). The $O_2(g)$ cycle – its atmospheric sources and sinks - could clearly be very different for Earth-like exoplanets. Potential sources associated with photosynthesis are likely linked to the net stellar energy reaching the surface but could also be regulated by geological processes including burial and plate tectonics. Potential $O_2(g)$ sinks could be controlled by the amount of reducing gases in the exoplanetary atmosphere which can be directly influenced by the stellar environment and by the interior via outgassing.

## 3.2. Ozone ($O_3$)
### 3.2.1 Modern Earth
**Introduction** - the ozone layer extends from about 20 to 50km with peak $O_3(g)$ mixing ratios of ~10 parts per million (ppm) occurring around 30km in the mid-stratosphere. Modern Earth's atmosphere contains in total ~$3x10^{15}$g $O_3(g)$. The overall column thickness of the ozone layer is frequently expressed in Dobson Units (DU) named after Gordon Dobson, a pioneer in atmospheric $O_3$ research. A typical mid-latitude $O_3(g)$ column value for modern Earth is 320 DU. Such a value means that if all $O_3(g)$ molecules in the atmospheric column were to be brought adiabatically down to sea level which is held constant at standard temperature and pressure, this would result in a compressed $O_3(g)$ surface layer 3.20mm thick. This illustrates the fragility of the $O_3(g)$ shield which protects all surface life on Earth by absorbing ~90% of the incoming, biologically-harmful UVB radiation. One DU corresponds to $2.69x10^{16}$ molecules $cm^{-2}$.

**Stratosphere** - Sydney Chapman first accounted for the existence of the ozone layer by proposing a sequence of gas-phase reactions (Chapman, [83]) which occur mainly in the stratosphere and are collectively known as the Chapman mechanism:

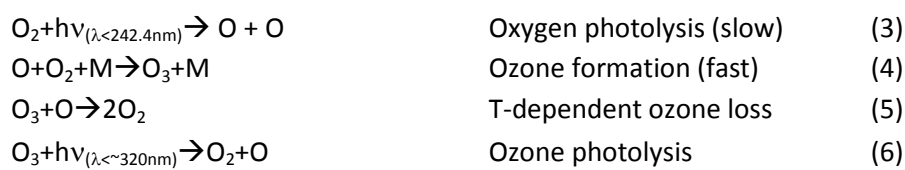

| | | |
|---|---|---|
| $O_2 + h\nu_{(\lambda<242.4nm)} \rightarrow O + O$ | Oxygen photolysis (slow) | (3) |
| $O + O_2 + M \rightarrow O_3 + M$ | Ozone formation (fast) | (4) |
| $O_3 + O \rightarrow 2O_2$ | T-dependent ozone loss | (5) |
| $O_3 + h\nu_{(\lambda<\sim320nm)} \rightarrow O_2 + O$ | Ozone photolysis | (6) |

Stratospheric $O_3(g)$ on Earth is therefore mainly formed from $O_2(g)$ in the presence of UV of the appropriate wavelengths (see e.g. Brasseur and Solomon, [33]) for a detailed discussion of the properties of Earth's ozone layer). $O_3(g)$ in Earth's atmosphere is thus a biosignature since it is formed chiefly from atmospheric $O_2(g)$ which in turn is formed mainly from life. In addition to being affected by the above four reactions, $O_3(g)$ is also destroyed in Earth's atmosphere by the so-called chemical "families" e.g. via HOx (in the stratosphere and mesosphere; Bates and Nicolet, [84]); ClOx (mainly in the upper stratosphere; Stolarski and Cicerone, [85]) and NOx (mainly in the lower stratosphere; Crutzen, [86]). Species are grouped into families in this way (e.g. HOx=OH+HO$_2$) because the individual family members (e.g. OH, HO$_2$ in the case of HOx) rapidly inter-react with each other depending on changing atmospheric conditions of e.g. p, T, and insolation. The essential, conserved quantity thereby is the sum of the inter-reacting species and it is therefore convenient to group such species together into families. The HOx, NOx and ClOx family members operate in catalytic cycles which efficiently destroy $O_3(g)$ as reviewed in Wayne [13]. An important generic cycle is as follows:

$$X+O_3 \rightarrow XO+O_2 \tag{7}$$
$$XO+O \rightarrow X+O_2 \tag{8}$$
$$\text{-----------------}$$
$$\text{net:} \quad O_3+O \rightarrow O_2+O_2 \tag{9}$$

where X=(e.g. Cl, OH, NO). The species 'X' which is consumed in reaction (7) is re-generated in reaction (8) and is therefore a catalyst which can participate in many such cycles. In this way, even trace amounts of 'X' (typically present at a few parts per billion by volume, ppbv) can overall destroy $O_3$ (reaction 9) which is present in much greater abundances (typically a few ppmv in Earth's stratosphere). Reservoir (or "inactive") molecules refer to atmospheric species which "store" HOx, ClOx and NOx but can release them depending on e.g. UV or temperature conditions. Important reservoirs in Earth's middle atmosphere include e.g. nitric acid ($HNO_3$) (for HOx and NOx) and hydrochloric acid (HCl) (for ClOx). Including the catalytic cycles in atmospheric models leads to a lowering in the steady-state solution for $O_3(g)$ by a factor of x2 to x3. The catalytic cycles are therefore clearly required in order to bring the modelled $O_3(g)$ abundance into line with observations.

In the lower stratosphere ozone has a rather long chemical lifetime of several weeks hence its abundance here is controlled mainly by transport ('dynamically-controlled'). In the upper stratosphere it has a shorter lifetime of the order of minutes to hours and is therefore mainly affected by chemistry ('photochemically controlled') (World Meteorological Organization (WMO) Report, [87]). Ozone is mainly formed in the tropics via the Chapman mechanism and is then transported to higher latitudes (where it is has its maximum column values) by the Earth's so-called Brewer-Dobson circulation in the stratosphere which moves air parcels upwards and polewards at the equator and downwards at higher latitudes as discussed e.g. in Stordal and Hov [88]. Ozone is a radiatively-active gas which heats the stratosphere and is responsible for a temperature rise with increasing altitude from ~20-30km. Ozone features a seasonal cycle of amplitude ~50DU with a late winter maximum due to suppressed photolytic loss.

In the lower stratosphere at high latitudes in winter/spring $O_3(g)$ is destroyed by catalytic cycles in the lower stratosphere involving ClOx. The phenomenon is popularly referred to as the "ozone hole". Its mechanism is related to the formation of polar stratospheric clouds at cold temperatures. On the surfaces of these clouds heterogeneous chemical reactions lead to the conversion of reservoir chlorine compounds (such as HCl) into active ClOx (such as Cl and ClO) which destroy $O_3(g)$ catalytically in the presence of sunlight (see e.g. the WMO report [87] and references therein). There is a rich literature of 3D model studies with coupled photochemistry-climate for the modern Earth (see e.g. the Intergovernmental Panel on Climate Change (IPCC) fourth assessment report [89] for an overview.

**Troposphere** – in the Earth's lowermost region (from ~0 to 20km) the Chapman mechanism slows considerably since the UV radiation needed to break $O_2$ (reaction 3) is weak. A different mechanism, sometimes termed the "smog mechanism" mainly produces ozone in the troposphere (Haagen-Smit, [90]). The smog mechanism requires so-called Volatile Organic Compounds (VOCs) (such as $CH_4$) in the presence of UV and is catalyzed by NOx. The main steps (with $CH_4$ as the VOC) are illustrated below. Species shown in grey occur on both sides of the chemical reaction system so may be omitted when deriving the net equation:

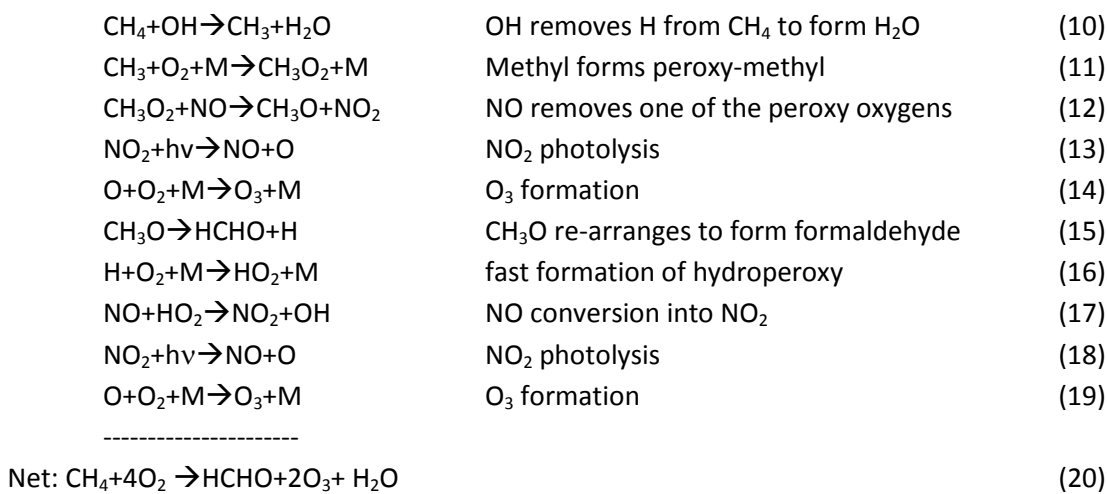

| | | |
|---|---|---|
| $CH_4+OH \rightarrow CH_3+H_2O$ | OH removes H from $CH_4$ to form $H_2O$ | (10) |
| $CH_3+O_2+M \rightarrow CH_3O_2+M$ | Methyl forms peroxy-methyl | (11) |
| $CH_3O_2+NO \rightarrow CH_3O+NO_2$ | NO removes one of the peroxy oxygens | (12) |
| $NO_2+hv \rightarrow NO+O$ | $NO_2$ photolysis | (13) |
| $O+O_2+M \rightarrow O_3+M$ | $O_3$ formation | (14) |
| $CH_3O \rightarrow HCHO+H$ | $CH_3O$ re-arranges to form formaldehyde | (15) |
| $H+O_2+M \rightarrow HO_2+M$ | fast formation of hydroperoxy | (16) |
| $NO+HO_2 \rightarrow NO_2+OH$ | NO conversion into $NO_2$ | (17) |
| $NO_2+hv \rightarrow NO+O$ | $NO_2$ photolysis | (18) |
| $O+O_2+M \rightarrow O_3+M$ | $O_3$ formation | (19) |

----------------------

Net: $CH_4+4O_2 \rightarrow HCHO+2O_3+ H_2O$                          (20)

In reaction (20) $CH_4$(g) is partially oxidized into formaldehyde, HCHO(g). This species however, can itself take part in further smog cycles where it replaces $CH_4$ as the VOC species. Thus the oxidation process continues and $CH_4$ is eventually fully oxidized into $CO_2$ plus $H_2O$ via the net reaction:

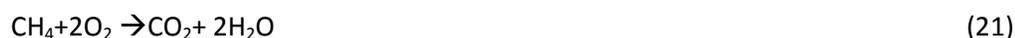

$CH_4+2O_2 \rightarrow CO_2+ 2H_2O$                                   (21)

On Earth the smog cycle can be limited by the concentration of either the VOC or NOx depending on conditions of T, p, UV, composition etc. Smog (tropospheric) $O_3$(g) typically makes up ~10% of the overall ozone column.

**Mesosphere** – above ~50km in the mesosphere, $O_3$(g) is destroyed directly via photolysis and also by catalytic HOx cycles which are driven in this region e.g. by water photolysis. Water abundances in turn are driven by photochemistry, diffusion and dynamical processes - so there is a link between mesospheric $O_3$(g) and atmospheric transport. $O_3$(g) features a secondary maximum in Earth's upper mesosphere. This could be linked with low temperatures there which slow ozone's photochemical sinks (see Smith and Marsh, [91] for a discussion).

### 3.2.2 Early Earth

The Earth's $O_3$(g) layer is thought to have mostly arisen almost simultaneously with the GOE (see e.g. discussion in Kasting and Catling [28]). Model studies e.g. Segura et al. [70]; Gebauer et al. [45] suggest that the $O_3$(g) layer reaches >90% of its modern value shortly after the GOE ~2.4 GYrs ago. The abundance of $O_3$(g) is generally rather stable against changes in its precursor $O_2$(g) due to the so-called ozone-UV (negative) feedback which operates as follows: consider an atmosphere in which the $O_2$(g) concentration decreases (increases). Then, the $O_3$(g) also decreases (increases) because its precursor $O_2$(g) decreases (increases) via the Chapman mechanism (see reactions 3 and 4 above). As a result, the incoming UV can now penetrate deeper into the atmosphere. This leads to an increase UV on the lower levels which can photolyse more $O_2$(g) (reaction 3) to form $O_3$(g) (reaction 4). The net effect is that the ozone peak moves down but overall its column value stays approximately constant.

### 3.2.3 Solar System

The photolysis of $CO_2(g)$ as discussed above can lead to formation of $O_2(g)$ and $O_3(g)$ abiotically. $O_3(g)$ has indeed been detected in the $CO_2(g)$-dominated atmospheres of both Venus and Mars, albeit in quite low amounts which we will now briefly discuss.

**Venus** - a tenuous, high-altitude (~100km peak) ozone layer has been detected in the Venusian atmosphere with values peaking at ($10^7$-$10^8$) molecules $cm^{-3}$ (Montmessin et al., [92]). This corresponds to a column value of ~(1/150)DU (=~$6.7 \times 10^{-3}$DU) which is clearly very thin compared with Earth's ozone column of ~320DU.

**Mars** – the (variable) ozone layer on Mars is linked with HOx abundances which control the re-generation of $CO_2(g)$ from its photolysis products (CO+O). HOx depends on water which is variable in time and space (see e.g. Yung and DeMore [36]). The Martian atmosphere features (0.3-3.0) DU which is (100-1000) times less than that of the Earth. The ozone layer on Mars has two maxima which both have abundances of up to ~$10^9$ molecules $cm^{-3}$ and which peak near the surface and at ~40km (Lefèvre et al., [93]). Their study suggested that the double maxima in the Martian ozone layer arise due to a complex seasonal behavior in water hence HOx which is a major ozone sink. Krasnopolsky [14]) reviews ozone observations in the atmosphere of Mars (and Venus). Several groups have applied 3D models studying the global ozone responses on Mars (e.g. Lefèvre et al., [94]; Sonneman et al., [95]; Moudden and McConnell, [96]).

**Icy Moons** – thin, $O_3(g)$ signals have been measured in the exo-atmospheres of some of the Galilean moons such as Ganymede (Noll et al., [97]) and Saturnian moons such as Rhea and Dione (Noll et al., [98]). The $O_3(g)$ is suggested to be formed due to the action of high-energy incoming ions upon oxygen-containing ices on the surface. Column ozone amounts typically range from several tenths up to about 2 DU.

### 3.2.4 Earth-like Exoplanets

A common modeling approach is to assume the Earth's size, development, biomass etc. then to perform a parameter study varying e.g. star-planet orbit, stellar spectrum, biomass emissions etc. to determine the effect upon habitability. The main advantages of this approach are: (1) it is based on planet Earth, where we know life developed, and (2) the models are being applied in a range where they are well-validated based on a relatively rich range of observations. A criticism of the approach however, is its Earth-centricity. Numerous model studies have adopted this approach to calculate atmospheric climate, composition and spectra of an Earth-like planet (see e.g. Segura et al., [70],[71]; Tinetti et al., [99]; Grenfell et al., [100]; Rauer et al., [5]; Hedelt et al., [101]; Rugheimer et al., [102],[103]). This indicates whether the planet is situated in the habitable zone and estimates the potential range of atmospheric spectral signals (see also Seager, [104]). A main focus is Earth-like planets orbiting cooler (e.g. M-dwarf stars) (see e.g. overview in Kasting et al., [105]). For such objects, modeling studies have investigated the effects of cosmic rays (Grenfell et al., [106]; Tabataba-Vakili et al., [107]) and flares (Segura et al., [108]) from the central star upon the planet's atmosphere, concluding that these can have a potentially important effect upon atmospheric biosignatures such as $O_3$(g). An important result of the above studies is that the incoming stellar spectrum is potentially a key parameter influencing planetary temperature and composition hence biosignature spectra.

In a different, less Earth-centric approach, models are applied to determine the central factors affecting planetary habitability. Raymond et al. [109] for example, investigated water delivery and suggested that the Earth received only modest amounts of water compared to other Earth-like planets simulated. Other potentially important factors include the presence of a large moon (Laskar et al., [110]) or the role of gas giants in the outer system (Horner and Jones, [111]).  Their exact role upon the habitability of Earth-like planets is however still debated.

### 3.2.5 Earth-like Exoplanets in time

Many processes (e.g. outgassing, escape, delivery etc.) can affect the evolution of a terrestrial planet's atmosphere into a habitable state. Lammer et al. [5] review the evolution of atmospheric habitability for terrestrial exoplanets. Regarding atmospheric biosignatures, a key event is the establishment of an $O_3$(g) layer since this protects the surface from UV which enables other biosignatures (such as $N_2O$(g) and $CH_3Cl$(g)) to build-up in the atmosphere (see e.g. Grenfell et al., [112]). Pilcher [113] discuss spectral biosignatures of "Early Earths" and noted the potential importance of organosulfur species. Des Marais et al., [75] and Kaltenegger et al. [114] discuss biosignature spectral signatures (including those for $O_3$(g)) over geological epochs for Earth-like planets.  Rugheimer et al. [102] modelled the evolution of atmospheric ozone columns for terrestrial planets orbiting FGKM stars (assuming different atmospheric compositions) at different geological epochs. Results suggested that pre-biotic Earth-like atmospheres receive up to x6 more (x300 less) biologically-active radiation when orbiting in the HZ of an F0V-star (M3.5-star) respectively.

### 3.2.6 Biosignature aspects

Ozone absorbs in the IR e.g. at 9.6 and 4.8 microns. The depth of the spectral bands is related to the temperature difference between the (warm on Earth) troposphere and (cool) stratosphere. This temperature difference is likely to be sensitive to the stellar class of the central star (see e.g. Rauer et al., [19] for a discussion). The so-called triple-signature (Selsis et al., [115]) hypothesis proposed that the simultaneous detection of atmospheric $O_3$(g), $CO_2$(g) and $H_2O$(g) bands is a more reliable biosignature than $O_3$(g) alone. This is because, high $CO_2$(g) abundances can mask the $O_3$(g) band and high $H_2O$(g) produces HOx which destroys $O_3$(g).  Thus, such an $O_3$(g)  band is more likely to arise biologically (since low $CO_2$(g) atmospheres have low $O_3$(g) abiotic production) but would nevertheless need to have a strong $O_2$(g) source to overcome catalytic loss from HOx.

Ozone retrieval (von Paris et al., [116]; Irwin et al., [117]) and detection (Hedelt et al., [101]) is very challenging - even for nearby (~10 parsecs) hypothetical Earth-like planetary targets using next generation instruments. Barstow et al. [118] suggested that atmospheric $O_3$(g) (assuming Earth's amount of this species) could be detected by averaging 30 transits observed by the James Web Space Telescope for the Earth-sized exoplanets TRAPPIST-1c and 1d. Other challenges include potential overlap of $O_3$(g) bands by $CO_2$(g) (Selsis et al., [115]; von Paris et al., [119]), dampening of the $O_3$(g) 9.6 micron band by the presence of clouds (Kitzmann et al., [120]) and potential interference of the $O_3$(g) band by the presence of a moon (Robinson et al., [121]). Barstow et al. [122] compare spectral retrieval (of e.g. $O_3(g)$) in the atmospheres of "exo-Earths" and "exo-Venuses".

### 3.2.7 Concluding remarks

Atmospheric ozone features fascinating interactions with biology, photochemistry, climate and transport. The presence of an ozone layer protects the lower atmosphere from UV which facilitates the presence of other atmospheric biosignatures such as $N_2O(g)$ which are destroyed by UV. $O_3(g)$ is a good indicator of $O_2(g)$ in the sense that it is formed primarily from $O_2(g$ and features a strong absorption band which is present over a large $O_2(g)$ range of several orders of magnitude. Drawbacks of interpreting $O_3(g)$ as a biosignature include overlap with $CO_2(g)$ and weakening of spectral bands due to clouds. $O_3(g)$ is produced mainly (~90%) via the Chapman mechanism on Earth but on exoplanets with weaker UV, $O_3(g)$ production via the smog mechanism (Grenfell et al., [123]) could be more important. Chemical families such as HOx, NOx, ClOx can efficiently reduce $O_3$ (by ~x2 to x3 on Earth). In an exoplanet context, the effect of HOx cycles for example could be stronger in in damp atmospheres, or worlds with high cosmic ray or/and lightning input (more NOx) or volcanically-active worlds (more chlorine outgassed hence more ClOx). The effect of interactions between the different families over the relevant parameter range is not well determined and their combined effect upon e,g, the false positives for $O_2$ and $O_3$ mentioned above - is an important area of future work,    Cases in which life is present but difficult to detect e.g. due to strong catalytic removal of $O_3$ - would represent "false negatives" when searching for life. Finally, $O_3(g)$ can be produced abiotically via the photolysis of $CO_2(g)$ (as on Venus and Mars); photolysis of $H_2O(g)$ followed by H-escape or, by the action of high energy ions on icy surfaces (as on the moons Ganymede, Rhea and Dione). This illustrates that the presence of $O_3(g)$ is not unambiguous and requires knowledge of the planetary environment when assessing as a biosignature.

### 3.3 Nitrous oxide ($N_2O$)
### 3.3.1 Modern Earth

$N_2O(g)$ (commonly known as "laughing gas") has been applied in medical science as an anesthetic and as an aerosol spray for cardiac patients. Earth's present-day atmosphere features an abundance, $\xi_{N2O}$=3.3x10$^{-7}$vmr (IPCC, [89]) which corresponds to a total mass in Earth's atmosphere of ~2.6x10$^{15}$g. The main $N_2O(g)$ atmospheric source arises via biogenic input from (de)nitrifying bacteria present in soils and sediments as part of the nitrogen cycle. The main $N_2O(g)$ atmospheric sinks (McElroy and McConnell, [124]) are:

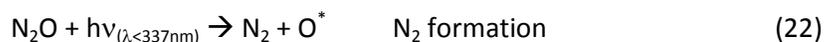
$$N_2O + h\nu_{(\lambda<337nm)} \rightarrow N_2 + O^* \qquad N_2 \text{ formation} \qquad (22)$$
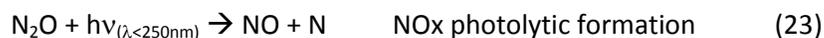
$$N_2O + h\nu_{(\lambda<250nm)} \rightarrow NO + N \qquad NOx \text{ photolytic formation} \qquad (23)$$
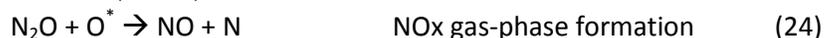
$$N_2O + O^* \rightarrow NO + N \qquad NOx \text{ gas-phase formation} \qquad (24)$$

where the asterisk indicates electronic-excitation. Syakila and Kroeze [125] review $N_2O(g)$ global atmospheric sources and sinks on the modern Earth. $N_2O(g)$ in today's atmosphere is mostly produced via biology. Abiotic $N_2O(g)$ production rates on present-day Earth are weaker than the estimated biotic rates by ~2 orders of magnitude. Levine et al. [126] suggested that lightning on Earth (and Venus) could be a weak, abiotic $N_2O(g)$ source. Samarkin et al. [127] noted an abiotic production mechanism for $N_2O(g)$ via reaction of aqueous nitrite with iron (II) minerals from the "Don Juan" Pond in Antarctica.

$N_2O(g)$ is an efficient greenhouse gas and is chemically rather unreactive in the troposphere. It therefore has a long atmospheric lifetime of up to several hundred Earth years and is a dynamical tracer of atmospheric motions (IPCC, [89]).

### 3.3.2 Early Earth

Buick [128] suggested that $N_2O(g)$ could have been a major greenhouse gas on the Early Earth, emitted from the strongly reducing, so-called "Canfield ocean". In this scenario the sulphur-rich early ocean impacted the availability of trace metals in the ocean which in turn affected the nitrogen cycle and enhanced biological production of $N_2O(g)$. Grenfell et al. [129] and Roberson et al. [130] performed model studies which suggested that $N_2O(g)$ could play a role in warming the Early Earth. Airapetian et al. [131] suggested that high energy particles from superflares from the young Sun could have led to abiotic fixation of $N_2(g)$ hence to efficient abiotic $N_2O(g)$ formation in the Early Earth's atmosphere.

### 3.3.3 Solar System

Earth is the only planet known to date with a dominantly biogenic source of $N_2O(g)$. Abiotic $N_2O(g)$ sources via coronal discharge have been suggested to occur in the atmospheres of early Venus and Mars (Mvondo et al., [132]) and on early Earth (Mvondo et al., [133]).

### 3.3.4 Earth-like Exoplanets

For Earth-like planets orbiting in the HZ of cooler stars, the low UV output from the star weakens $N_2O(g)$ sinks (e.g. reactions (22, 23)) hence favors its build-up in the atmosphere. Modeling studies (e.g. Segura et al., [72]; Grenfell et al., [112]; Rugheimer et al., [103]) have suggested this effect. Since $N_2O(g)$ is expected to a long lifetime in Earth-like atmospheres its abundance could be controlled by atmospheric dynamics which is not well-constrained on such worlds. Studying the interactions of $N_2O(g)$ atmospheric photochemistry and dynamics therefore represents important future work for 3D Earth-like modeling studies.

### 3.3.5 Earth-like Exoplanets in time

Superflares (Airapetian et al., [131]), see above) could lead to abiotic $N_2O(g)$ production in $N_2$-$O_2$ atmospheres. This process could therefore be relevant for Earth-like planets orbiting younger stars and for planets in the close HZ of active cooler stars and needs further investigation. This suggests a trade-off between on the one hand the superflare breaking $N_2(g)$ hence forming $N_2O(g)$ and on the other hand high UV destroying $N_2O(g)$ photolytically.

### 3.3.6 Biosignature aspects

$N_2O(g)$ bands occur in Earth's infrared atmospheric windows, e.g. at 7.8, 4.5 and the weaker $3.7\mu m$ bands (Muller, [134]). For Earth-like planets these bands are relatively weak compared with other biosignature (e.g. $O_3(g)$) features but they could become stronger e.g. for planets orbiting stars with weak UV-output since UV is an important sink for $N_2O(g)$ (e.g. Segura et al., [72]). $N_2O(g)$ spectral features also become more apparent for Earth-like planets with lowered $CH_4(g)$ emissions (Grenfell et al., [112]) which was attributed to a cooling effect in the middle atmosphere.

### 3.3.7 Concluding remarks

$N_2O(g)$ sources on modern Earth are dominantly due to biology. In an exoplanetary context there could be abiotic sources under strongly flaring conditions which need further investigation. $N_2O(g)$ is an important source of NOx source in Earth's stratosphere; this represents a potential coupling between $N_2O(g)$ and $O_3(g)$ because NOx is an efficient ozone sink in the stratosphere. NOx can however represent a source for $O_3(g)$ via the smog mechanism in the troposphere.

### 3.4 Methane ($CH_4$)

### 3.4.1 Modern Earth

Earth's present-day atmosphere features a methane abundance, $\xi_{CH4}=1.8\times10^{-6}$ vmr which corresponds to a total mass of ~$5.1\times10^{15}$g. The main surface source of $CH_4(g)$ (~500Tg/year) into the atmosphere arises due to a net flux from biology. A smaller source of a few tens of Tg/yr arises from geological processes (Bousquet et al., [135]) including serpentinization (i.e. the process of mineral alteration due to introduction of water into the rock system). Further understanding $CH_4(g)$ sources due to serpentinization is an important area of future study in the ocean community. On modern Earth, atmospheric methane has approximately tripled since the start of the industrial revolution at ~1850, mainly due to animal husbandry and agriculture. The main atmospheric $CH_4(g)$ sink is due to reaction with the hydroxyl (OH) radical (see above) in the troposphere with minor sinks due to reaction with soils, photolysis and gas-phase reactions e.g. with chlorine atoms or with electronically-excited oxygen atoms. $CH_4(g)$ displays inter-hemispheric and annual cycles with a magnitude of a few percent (see e.g. Saad et al., [136]) the causes of which are not well-constrained. $CH_4(g)$ has a rather long atmospheric lifetime of ~8 years (IPCC, [89]) and is therefore a dynamical tracer of motions in Earth's atmosphere (as is $N_2O(g)$, see 3.3.1).

### 3.4.2 Early Earth

Model studies proposed high $CH_4(g)$ values of (100-1000) ppmv (e.g. Catling et al., [137]) on the Early Earth compared with 1.8ppmv today. The high values were likely associated with strong net biological fluxes related to the reducing conditions in Earth's early environment. Abiotic $CH_4(g)$ sources were likely stronger too than today e.g. due to enhanced volcanic activity (Kasting and Catling, [28]). The response of the main $CH_4(g)$ sink due to reaction with OH is rather complex and depends on opposing processes. On the one hand higher UV output of the early Sun favors OH production (reaction 1). On the other hand, a fainter sun can lead to cool, dry conditions which can weaken OH production (reaction 2). Enhanced greenhouse warming from high $CH_4(g)$ on the Early Earth has been proposed as a possible solution to the Faint Young Sun paradox (see e.g. Catling et al., [137]).

### 3.4.3 Solar System

On Mars, observations by Formisano et al. [138] suggested a $CH_4(g)$ mixing ratio, $\xi_{CH4}=10\pm5$ppbv in the atmosphere. Krasnopolsky et al. [139] calculated a rather long $CH_4(g)$ lifetime against photochemical loss, $\tau_{CH4}$~340 Earth years (which suggests $CH_4(g)$ to be uniformly mixed in Mars' atmosphere). Subsequent data (Mumma et al., [140]) however, noted localized, non-uniform abundances in the form of plumes. This suggested a much lower $CH_4(g)$ lifetime compared with that on Earth of (0.4-60) Earth years (with a low, global mean abundance, $\xi_{CH4}$ of ~2ppbv). Their study suggested that heterogeneous chemistry in Mars' highly-oxidizing, dusty environment could lead to oxidant-coated particles lifted in the atmosphere which could efficiently remove $CH_4(g)$ hence lower its atmospheric lifetime. Lefèvre and Forget [141] discussed the discrepancy in modeled and observed $CH_4(g)$ lifetimes in a global context. Knak Jensen et al. [142] suggested that eroded mineral grains in the atmosphere provide the missing $CH_4(g)$ atmospheric sink by providing active sites for heterogeneous $CH_4(g)$ removal by forming $Si-CH_3$ bonds.

Could the reported $CH_4(g)$ signals on Mars represent an atmospheric biosignature? Atreya et al. [15] review possible sources, including biology geology and meteorites. Zahnle et al. [143] revisited the preceding $CH_4(g)$ data. Their analysis suggested $\xi_{CH4}$<3ppbv on Mars and noted the challenge of removing the imprint of the Earth's atmosphere. Recent in-situ data however, has increased confidence that $CH_4(g)$ is indeed present in the Martian atmosphere - Webster et al. [144] performed spectroscopic measurements with the Curiosity Rover at the Gale Crater which suggested variation mostly in the nominal range of (0.7-2.1)ppbv $CH_4(g)$ over a period of 605 Sols. The data also suggested controversial, temporal spikes of up to ~7ppbv $CH_4(g)$, the origin of which is hotly debated.

$CH_4(g)$ is also present in outer solar system bodies such as Titan, but is generally not considered to be a biosignature. This is because in these regions, carbon is (mainly) naturally present in reduced forms such (see Öberg et al., [145]). Lunine and Atreya [146] review the methane cycle on Titan.

### 3.4.4 Earth-like exoplanets and their evolution

Numerous modeling studies (e.g. Segura et al., [72]; Rauer et al., [19]; Grenfell et al., [112]; Rugheimer et al., [103] calculated enhanced (x100-x1000) atmospheric $CH_4(g)$ compared with modern Earth for Earth-like planets orbiting in the HZ of M-dwarf stars. This result is generally attributed to the weaker UV-output from the central star which leads to a reduction in $OH(g)$ (reactions 1 and 2), the main sink for $CH_4(g)$ in the planet's atmosphere. The range of abiotic $CH_4(g)$ production via geological processes on Earth-like planets was estimated by Guzmán-Marmolejo et al. [147]. Their results suggested that $N_2-O_2$ atmospheres with $CH_4(g)$ mixing ratios, $\xi_{CH4}$>10ppmv could not be produced abiotically and was therefore an indication of life. Schindler and Kasting [75] suggested that $CH_4(g)$ could be observable by next generation instruments from space for Earth-like planets with >100ppm. Arney et al., [148] suggested that production of organic hazes similar to those on the Archaean Earth could be an atmospheric biosignature in Earth-like atmospheres since such hazes require more $CH_4(g)$ to form than could be delivered by abiotic processes alone.

### 3.4.5 Biosignature aspects

$CH_4(g)$ features absorption bands at ~3.4μm and ~7.7μm (e.g. Rauer et al., [5]; Werner et al., [149]) which are blended with $H_2O(g)$ absorption bands at low resolution (R~20) for Earth-like atmospheres (Pilcher, [113]). Bands at ~1.7μm and ~2.4μm also become evident for $CH_4(g)$>100ppm (Des Marais et al., [77]).

**3.4.6 Concluding remarks**

Detection of atmospheric $CH_4(g)$ in Earth-like atmospheres is a possible indication of life but requires further information to exclude an abiotic origin such as volcanism (see Kaltenegger et al., [76]). There are interactions between $CH_4(g)$ and atmospheric $O_3(g)$. $CH_4(g)$ heats the middle atmosphere which can decrease $O_3(g)$ by increasing the rate of reaction 5. Also it is oxidized into water in the stratosphere which can form HOx hence destroy $O_3(g)$ catalytically. Increased $CH_4(g)$ could however lead to enhanced ozone formation by stimulating the initial step of the smog mechanism (reaction 10) if conditions are VOC-controlled. Further investigations are required to elucidate these interactions.

The four atmospheric species ($O_2$, $O_3$, $N_2O$, and $CH_4$) mentioned so far are generally the most discussed in the atmospheric biosignature literaure. Now, we provide a brief overview of other atmospheric species which have been considered in a biosignature context.

**3.5 Additional Atmospheric Biosignatures**
**3.5.1 Sulfur-containing gases**

Domagal-Goldman et al. [150] and Pilcher [113] discussed multiple sulfur-containing compounds (such as $CH_3SCH_3$, $CH_3S_2CH_3$) as potential atmospheric biosignatures in Earth-type environments with low-UV conditions e.g. for planets orbiting in the HZ of cool M-dwarf stars. The Domagal-Goldman et al. [149] study suggested that although organic sulfur-containing gases did not build up to levels which are detectable for next generation missions, their presence could nevertheless be indirectly elucidated since they led to an increase in the ($C_2H6/CH_4$) ratio. Seinfeld and Pandis [151] (their Table 2.3) give an overview of atmospheric budgets and lifetimes of sulfur-containing species in Earth's atmosphere.

**3.5.2 Chloromethane**

The modern Earth features a chloromethane ($CH_3Cl(g)$ abundance of ~0.6ppb with a lifetime of 1-2 years (IPCC, [89]). The global budget is not well-constrained. Biological sources include ocean plankton, fungi and wood rotting whereas important sinks are reaction with OH(g) and biological degradation (Harper, [152]; Keppler et al. [153]). Abiotic sources include methylation of chloride (Keppler et al., [153]). $CH_3Cl(g)$ has been assessed as an atmospheric biosignature on Earth-like planets by e.g. Segura et al. [72] and Grenfell et al. [112]. Spectral bands are rather weak (e.g. at 13.7 microns) although these could become enhanced for Earth-like plants orbiting cool M-dwarf stars (Segura et al., [72]; Rauer et al., [19]).

**3.5.3 Non-Methane Hydrocarbons**

Ethane ($C_2H_6(g)$) has been suggested as an exoplanetary atmospheric biosignature on Earth-like terrestrial planets (Domagal-Goldman et al., [150]). On modern Earth it has a concentration of a few ppbv and exists mainly in the lower troposphere. Its main sources (estimated to be 8-18Tg/yr) are anthropogenic from e.g. fossil fuel burning and it is removed mainly via reaction with OH (see Xiao et al., 154] for a review).

Isoprene ($C_5H_8$) (or 2-methyl-1,3-butadiene, $CH_2=C(CH_3)-CH=CH_2$) is commonly emitted by plants. Its global source of ~500Tg/yr (Seinfeld and Pandis, [151]) on modern Earth is comparable with methane but its concentration (a few ppb) is much lower since its lifetime against atmospheric destruction (mainly via reaction with OH) is short (~1 hour). Isoprene emissions are sensitive to temperature and light. Isoprene and its photochemical products can maybe influence aerosol formation although the mechanism is not well understood. Although isoprene is sometimes mentioned as a potential biosignature in the exoplanet literature (e.g. Seager et al., [12]) it is not widely studied in this context e.g. due to its rather fast atmospheric sinks (hence low abundance) compared with other proposed gas-phase biosignatures. Shaw et al. [155] review the atmospheric budget of isoprene on modern Earth.

### 3.5.4 Technosignatures

Potential signs of advanced life are sometimes termed "technosignatures". Lin et al. [156] considered potential atmospheric spectral signals from pollutants such as chlorofluorocarbons in Earth-like atmospheres. Stevens et al. [157] discussed possible left-over signatures of civilizations which have become extinct e.g. due to nuclear war, atmospheric pollution etc. Griffith et al. [158] and Korpela et al. [159] discuss potential technosignatures.

### 3.5.5 Atmospheric Redox Disequilibrium

Lovelock [160] and Lederberg [161] suggested that the simultaneous presence of significant amounts of oxidizing species (such as $O_2$) and reducing species (such as $CH_4$) in Earth's atmosphere could be interpreted as a biosignature. These two species have strong biotic sources via cyano- and methanogenic bacteria respectively. Without life, the atmospheric concentrations of strongly reducing and oxidizing species would be significantly lowered (in the case of $CH_4$-$O_2$, within several thousand years i.e. relatively quickly in terms of geological timescales). Note that the false positive mechanisms for generating $O_2$(g) discussed above (section 3.1.4) all involve a distinct absence of $CH_4$(g) i.e. consistent with the chemical disequilibrium framework.

Sagan et al. [162] suggested that simultaneous measurements of $CH_4$(g) and $O_2$(g) in Earth's atmosphere made by the Galileo spacecraft are tell-tale signs of life on our planet. Simoncini et al. [163] estimated with a chemical model that ~0.67 terawatts of power are supplied by life on Earth in order to maintain this $CH_4$-$O_2$ chemical disequilibrium. Court and Sephton [164] noted that methane sources from ablating carbonaceous micrometeorides could act as a false positive for ($O_2$-$CH_4$) biosignature candidates in some cases. Calculations by Krissansen-Totton et al. [165] suggested that chemical disequilibrium in Earth-like atmospheres could arise due to the presence of abundant gas-phase ($N_2$-$O_2$) together with liquid water. Without life, these species would form the stable nitrate ion in the ocean. Stüeken et al. [166] suggested that an atmosphere rich in both $N_2$(g) and $O_2$(g) indicates the presence of an oxygen-producing biosphere. Rein et al., [167] cautioned that the interfering presence of a moon with its own atmosphere could behave as a "false positive" for redox disequilibrium in the planetary atmosphere by contaminating biosignature band(s).

### 3.5.6 Isotope Ratios of Atmospheric Species

Life tends to favor the lighter isotope (see e.g. Horita, [168]) for its enzyme-driven biochemistry which is often highly specific. Detecting isotope ratios which are displaced towards the lighter element compared with a pre-defined, abiotic benchmark could therefore represent a sign of life. The measurement is however very challenging and generally not widely studied in the exoplanetary biosignature literature. Snellen [81] suggested that the proposed European Extra Large Telescope (E-ELT) could possibly distinguish isotope ratios in exoplanetary atmospheres (although he did not specify for which types of planets or atmospheric species). Yan et al., [169] suggested they could distinguish between modern Earth's atmospheric oxygen isotopes based on Earth-shine using a ground-based high resolution spectrograph. Regarding carbon isotopes, photosynthesis preferentially extracts the lighter isotope from the atmosphere although the effect is estimated to be small - of the order -25‰ (parts per thousand) fractionation in $^{13}C$ (see e.g. Kump, [170]; Holmen [171]). Hedelt et al. [172] showed that it is possible to distinguish atmospheric isotopes in the atmosphere of Venus using Earth-based IR spectroscopy from the ground. In summary, although an approach based on isotopes, especially carbon isotopes, offers a potentially powerful method to distinguish between biological and abiotic processes, their detectability is likely beyond the capability of next generation observatories.

## 4. Spectropolarimetry

Analysis of Earthshine (light reflected from the Earth to the moon and back) using spectropolarimetry e.g. by Sterzik et al. [21] suggests the presence of redox disequilibrium in Earth's atmosphere i.e. the simultaneous presence of $CH_4(g)$ and $O_2(g)$ at abundances which are un-explainable without invoking life. Hoeijmakers et al. [173] present a feasibility study for a spectropolarimeter placed on the moon to study Earth's polarization spectrum. Circular spectropolarimetry can in principle detect chiral signals e.g. due to chlorophyll, as recently discussed e.g. in Lucas Patty et al. [174] although its application as a biosignature beyond the Earth is very challenging.

## 5. Biochemical Network Properties

Biochemical reaction systems differ in their network properties compared with abiotic systems. For example, the likelihood of any given species behaving as either a reactant or a product is approximately equal for all species in biochemical reaction networks but this is generally not the case for abiotic networks (Jolley and Douglas, [175]). Constraining the amounts of relevant species in biochemical reaction networks is likely to be very challenging for Earth-like exoplanets. Walker and Davies [176] studied biochemical networks and modeled the effect of life upon so-called "transfer entropy" - the flow of information over spatial scales.

## 6. Abiotic Earth ('Dead Earth')

Margulis and Lovelock [177] estimated the change in Earth's atmospheric composition for a planet similar to Earth (in mass, radius, stellar environment etc.) - but without life. Such a world is sometimes referred to as an "Abiotic" or "Dead Earth". This is clearly important information in order to have a benchmark against which to compare when assessing biosignature candidates. $O_2(g)$ on a Dead Earth is removed e.g. via deposition and $N_2(g)$ is oxidized into NOx by lightning (see e.g. Tie et al., [178] and cosmic rays. The NOx produced is oxidized by OH into nitric acid ($HNO_3$) which is highly water-soluble hence efficiently rained out of the atmosphere in the form of the nitrate ($NO_3^-$) ion. Ultimately therefore, nitrogen from the atmosphere is converted into the stable, solvated nitrate ($NO_3^-$) anion in the ocean (as already discussed in section 7). The $CO_2(g)$ abundance in the atmosphere of a Dead Earth is not well-constrained. Margulis and Lovelock [177] suggested a range of (0.3-1000)mb surface $CO_2(g)$. Morrison and Owen [179] suggested however that all the Earth's present budget of $CO_2$ (~69 bar) could be returned to the atmosphere of a Dead Earth. These large uncertainties arise because the way in which life affects Earth's carbon-cycle (e.g. via the net removal of organic carbon in the ocean) is not well-known. O'Malley-James et al. [180] studied an Earth scenario where the biosphere declines and dies at ~2.8Gr in the future due to a brightening sun. Their results suggest that warmer, wetter climate first leads to enhanced carbon draw-down such that photosynthesis ceases when atmospheric $CO_2(g)$ falls below 10ppmv. Several works (e.g. Fraedrich et al., [181]; Kleidon and Fraedrich, [182]) applied climate models to study the influence of vegetation and surface-type upon planetary climate and hydrology.

## 7. Astrobiological Roadmaps and Biosignature Related Missions

The theme of exoplanetary biosignatures is firmly anchored in the current scientific roadmaps of ESA (see e.g. Horneck et al., [1]) and NASA. ESA's Cosmic Vision 2015–2025 program includes the key question: "How will the search for and study of exoplanets eventually lead to the detection of life outside Earth?" NASA's Astrobiology Institute Strategy Roadmap includes the theme: Identifying, exploring, and characterizing environments for habitability and biosignatures.

The ongoing Gaia mission (e.g. Prusti et al., [183]) will considerably advance our knowledge of stellar astrophysical properties for a large, representative sample of stars in the Galaxy. The Transiting Exoplanet Space Survey (TESS) (Ricker et al., [184]) and the CHaracterising ExOPlanets Satellite (CHEOPS) (Fortier et al., [185]) missions will improve understanding of hot Jupiters, mini gas planets and hot super-Earths. The PLATO 2.0 (Rauer et al., [186]) mission will characterize age and bulk distribution for a large number of rocky planets in the HZ - an important target pre-selection for follow-up missions searching for life. The James Webb Space Telescope (JWST) (Lightsey et al., [187]) will perform unrivalled atmospheric spectroscopy on mini gas planets and super-Earths. Nevertheless, detecting atmospheric biosignatures will be a major challenge for this mission. Léger et al. [188] suggested that the (still evolving) value of eta-Earth could influence the design of future space missions (e.g. via coronography or interferometry) searching for exoplanetary biosignatures. The next generation of large ground-based telescopes such as the European Extremely Large Telescope (e.g. Snellen, [81]) will perform high resolution spectroscopy of Terrestrial exoplanets from the ground.

In the further future, the Large Aperture UV-Optical-Infrared (LUVOIR) telescope is a space mission (Bolcar et al., [189]) searching for atmospheric biosignatures as part of NASA's 30-year roadmap. The Habitable Planet Imaging (HabEx) mission (Mennesson et al., [190]) is currently being considered by NASA with a suggested launch date in the 2030s. The mission plans to feature a >3.5m optical mirror in space performing direct imaging and spectroscopy to determine habitability and biosignatures. Mawet et al. [191] discuss technical advances in coronographic design. Turnbull et al. [192] discuss the starshade mission which consists of a proposed 4m telescope together with a 50m starshade, placed at the second Lagrangian (L2) point to characterize atmospheric spectral features such as $O_2(g)$ and $O_3(g)$ in Earth-like atmospheres. The Origins Space Telescope is the mission concept which arose from NASA's far infrared surveyor as part of the 2020 decadal survey. Proposed is a large aperture (>8m), cryogenically-cooled instrument (e.g. Meixner et al. [193]) which could therefore in theory perform transit spectroscopy on rocky exoplanets.

## 8. Summary

The search for atmospheric biosignatures is an integral part of the dedicated astrobiological roadmaps of ESA and NASA. Activities have recently gathered momentum due to the discovery of a planet orbiting in the potentially temperate zone around our nearest star, the red dwarf Proxima Centauri (Anglada-Escudé et al., [194]) which has since generated numerous modeling studies investigating its potential habitability.

There is an expanding literature which deals with exoplanetary atmospheric biosignatures (e.g. for $O_2$, $O_3$, $CH_4$, $N_2O$ etc.) including their potential responses and spectral signals. Recent years have seen a rather intense period of study of the potential abiotic sources of $O_2(g)$. This has led to a somewhat changing philosophy, namely that each biosignature must be considered in combination with the planetary and stellar environment.

The pathway to determining exoplanetary biosignatures is a challenging process. It involves establishing suitable targets of Earth-like planets which are habitable (see Williams and Gaidos, [195] who discuss the challenge of detecting liquid oceans). Nothing of course is known concerning which fraction of habitable planets (Cockell, [196]) could actually be inhabited. Such issues are, at least beginning to be addressed e.g. Lingam, [197] who attempt the formidable task of modeling the spreading of life via panspermia. The role of other factors which could affect the development of life e.g. planetary magnetic field or/and the existence of plate tectonics have, at least in theory suggested observables (see e.g. Parnell, [198]) although no data will be available in the near future.

There is an expanding literature of exoplanetary biosignatures which compiles theoretical spectra over a wide range of conditions. It is important is to distinguish the contribution of different atmospheric regions (e.g. mesosphere, stratosphere, troposphere) to the calculated spectroscopic signals (see e.g. Vasquez et al., [199,200]) as a function of composition, incoming insolation etc. This is because 1) lower layers below about 12-14km for Earth-like planets may not be sampled by transit spectroscopy due to atmospheric refraction (see e.g. García Muñoz et al., [201]; Bétrémieux and Kaltenegger, [202]), and (2) on Earth mesospheric (~70-80km) signals of some biosignatures (e.g. especially $O_2$, $O_3$) have very different sources and sinks compared with lower regions. Ongoing and planned missions in exoplanet science offer a diverse and exciting range of platforms with which to explore the fascinating question of whether life exists beyond the Solar System.